\documentstyle[12pt,prl,aps]{revtex}

\draft
\newcommand \be  {\begin{equation}}
\newcommand \bea {\begin{eqnarray} \nonumber }
\newcommand \ee  {\end{equation}}
\newcommand \eea {\end{eqnarray}}
 \def\(({\left(}
 \def\)){\right)}
\def\bi{\bibitem}
\def\s{\sigma}
\def\pesc{p_{esc}}
\begin{document}

\title{
Aging without disorder on long time scales}
\author{Werner Krauth$^{(1)}$ and Marc M\'{e}zard$^{(2)}$}
\address{
$^{(1)}$ CNRS-Laboratoire de Physique Statistique de l'ENS\\
24, rue Lhomond; F-75231 Paris Cedex 05; France\\
e-mail: krauth@physique.ens.fr\\
$^{(2)}$ CNRS-Laboratoire de Physique Th\'{e}orique de l'ENS$^*$\\
24, rue Lhomond; F-75231 Paris Cedex 05; France\\
e-mail: mezard@physique.ens.fr}
\date{July 1994}
\maketitle
\begin{abstract}
We study the Metropolis dynamics of a simple spin system
without disorder, which exhibits glassy dynamics
at low temperatures.
We use an implementation of the  algorithm  of Bortz, Kalos
and Lebowitz  \cite{bortz}.
This method turns out to be very efficient for
the study of  glassy systems, which get
trapped in local minima on many different time scales.
We find strong
evidence of aging effects at low temperatures. We
relate these effects to the
distribution function of the trapping times of
single configurations.

LPTENS preprint 94/20.
\end{abstract}
\vskip 10cm
$^*$ Unit\'e propre du CNRS, associ\'ee \`a l'ENS et \`a
l'Universit\'e de Paris XI
\newpage

In the last few years, a lot of work has been devoted to
the description of systems with glassy dynamics.
The phenomenon of  aging\cite{ageexp} seems to be ubiquitous.
Dynamical aging, which may be loosely
defined as the property that correlation or response functions
on a time scale $\tau$ depend on the age of the system
$t_w$ when $\tau \simeq t_w$, has been used as a powerful probe
in experimental studies of the spin glass phase \cite{agefra}.
Recently, a lot of numerical studies of various spin glass systems
have reported aging effects \cite{agesimu}. The theoretical understanding
of aging is still
not satisfactory, although there have been interesting
proposals of phenomenological models \cite{kophil,fishus,bouchaud},
and also
some attempts at a microscopic study and a link with the static
replica solution of some simple mean field systems
\cite{horner,dfi,cuku,framez,cukusk}.

In the spin glass case,  slow dynamics is generated
by the frozen  disorder. In this paper we want to report on the
numerical observation of aging effects in a model which
has no quenched disorder. This model is a simple spin system
with long-range four-spin interactions which has been
studied from a physical point of view by Bernasconi \cite{bernasconi}.
That quenched disorder is not mandatory for the existence
of a glassy dynamics is evident: the glassy state is a
notorious counter example. The system we study provides
a simple model for this effect.
In a sense, the system generates its own
disorder through the slowing down of some degrees of freedom.
This model can help to provide a new bridge  between
glassy systems with or without quenched disorder, on top of
the approaches through mode coupling theory \cite{gotze},
or some other glass models
studied recently \cite{golzip}. The emergence of long time scales
in frustrated systems without disorder has been noticed in recent papers
\cite{sethna}.

The numerical observation of some non-equilibrium
phenomena which resemble  aging
in a Monte Carlo simulation is rather easy. A more difficult
task is to assert whether these are just transient phenomena
taking place on time scales shorter than some equilibration time
$t_{eq}$ or if $t_{eq}$ is infinite.
This task is difficult almost by definition:
in the glass phase the dynamics is slow,
and aging should be demonstrated by the existence
of relaxation processes taking place on very
different time scales, in order to be
safely distinguished from a simple, but slow relaxation rate
towards equilibrium. This requires simulating a system
at low temperatures, in a regime where the acceptance rate
of the Metropolis algorithm becomes very small.

This problem of the low acceptance probability
can be circumvented efficiently with an
algorithm, originally proposed by Bortz {\sl et al} \cite{bortz},
and used in some rare occasions\cite{sethna} \cite{kinzel}.
We introduce here an efficient implementation of this algorithm,
and apply it to the Bernasconi model. At low temperatures, we can
simulate this system on time scales which
are several orders of magnitude larger than those accessible
by a straight Monte Carlo simulation.
We suspect
that this algorithm could be quite useful for the simulations
of a whole class of glassy systems at low temperatures, which
we shall characterize briefly in the conclusion.

The problem studied by Bernasconi some years ago
consists in finding  low autocorrelation binary
sequences (LABS).
This optimization problem, originally introduced for
its applications in communications,
can be turned into a physical problem of N Ising spins, where the $2^N$
spin
configurations are weighted by a Boltzmann factor,
with an energy given by:
\begin{equation}
{\cal H} = {1 \over N} \sum_{k=1}^{N-1} \left [\sum_{i=1}^{N-k} S_i
S_{i+k}\right]^2
\end{equation}
Monte Carlo simulations have been carried out by Bernasconi
\cite{bernasconi},
Golay \cite{golay},
and more recently by Migliorini \cite{miglio}, who have shown that
finding the ground state (or low lying configurations)
is numerically hard; the system freezes at a certain
temperature, below which the specific heat becomes
very small: the system gets trapped into metastable
states.
Here we would like to consider it as a simple prototype of
systems exhibiting dynamical aging and undergoing a glass transition.
Simultaneously to our numerical simulations, several independent
works have tried to
investigate this problem \cite{bm}, or another version with periodic
boundary conditions \cite{mpr}, analytically by approximating
it by  a disordered system. The corresponding disordered problem
can be studied by the replica technique and undergoes a "one step
replica symmetry breaking" phase transition characteristic of
a class of spin glass transitions of the category of the
Random Energy Model \cite{derr,gm}. Such a
freezing  scenario
is well confirmed by our simulations: at low temperatures,
the system gets trapped into single configurations which it
takes many Monte Carlo steps to escape from.

Let us now describe the algorithm we have been using.
We  adopt the usual single spin-flip dynamics,
which produces a sequence of configurations
\begin{equation}
\sigma_1(\tau_1) \rightarrow \sigma_2(\tau_2)
 \ldots \rightarrow \sigma_i(\tau_i)
 \rightarrow \sigma_{i+1}(\tau_{i+1})\ldots
{\sl etc}
\label{MCsequence}
\end{equation}
In eq. (\ref{MCsequence}),
the symbols indicate  that the system stays in configuration
$\sigma_i$ for a time $\tau_i$  etc.
The transition probability is non-zero only for configurations
differing by a single spin. (We denote in the following by
$\sigma^{[m]}$ the configuration which results from $\sigma$ after
flipping spin $m$). We use the Metropolis dynamics, in which
 the transition probability per unit time  $\Delta\tau$
 is given
by the explicit formula:
\begin{equation}
p(\sigma \rightarrow \sigma^{[m]}) = \frac{\Delta \tau}{N}
\left\{ \begin{array}{l}
1 \;\;\;\;\;\;\;\;\;\;\;\;\;\;\; (if\;\;\; E(\sigma^{[m]}) < E(\sigma)) \\
\exp[-\beta( E(\sigma^{[m]}) - E(\sigma))]\;\;\ (otherwise)
\end{array} \right. m=1,\ldots, N
\label{transition}
\end{equation}
Usually, the sequence in eq. (\ref{MCsequence})
is directly simulated
by means of
a rejection method. The move $\sigma \rightarrow \sigma^{[m]}$ is accepted
with probability eq. (\ref{transition}), otherwise the system stays in
configuration $\sigma$. The method
we use calculates the trapping times $\tau_i$ not by
repeatedly rejecting moves, but by sampling $\tau_i$ from its
probability distribution $P_\s(\tau)$; it then  cashes
 in repeatedly on the
initial expenditure
of calculating $P_\s(\tau)$.

When one reaches a configuration $\sigma$, the transition probabilities
$p_m= p(\sigma \rightarrow \sigma^{[m]})$
towards
{\it all} the neighboring configurations $\sigma^{[m]}, \  m=1,...,N$ are
calculated. The escape probability from configuration $\s$ is
$p_{esc}=\sum_m p_m$, and the probability of leaving $\s$ after
exactly $\tau$ MC is equal to $P_\s(\tau)=\pesc (1-\pesc)^{(\tau-1)}$.
The conditional probability for reaching configuration
$\sigma^{[m]}$, given that one leaves $\s$, is $q_m=p_m/\pesc$. From
$\s$, the algorithm samples the time $\tau$ at which
the next visited configuration will be reached, using the distribution
$P_\s(\tau)$.
It samples independantly the new configuration $\sigma^{[m]}$
(automatically different from $\s$)
 by choosing the value of $m$ with probability $q_m$.
This algorithm is exactly equivalent to the standard one.

 The price one pays in this  algorithm is to compute
for each visited configuration all the  $p_m,  \  m=1,...,N$, while
only one is computed in the usual method. Therefore this algorithm becomes
more efficient roughly speaking when the acceptance
rate of the usual rejection method becomes of order 1/N \cite{slow}.
Furthermore we have implemented in the algorithm an important feature
which provides a large gain in computer time, at least on the LABS
problem. We constantly
keep in memory an ``archive''  of the {\it different}
configurations visited recently
during the simulation and of the corresponding probabilities
$p_m $ for $m=1,...,N$.
Upon encountering  a configuration $\sigma$ which belongs to the
archive we use the $p_m $ in memory.
In our implementation the archive  is emptied
automatically as soon as the number of archived configurations reaches
a number
 $N_{arch}$. In the actual calculations described below,
for the runs at the smallest temperature $T=.075$, a configuration
which enters the archive (with $N_{arch}\sim 400$) typically remains
active for a very long time (during $\sim 10^6$ configurations), before
it is moved out.

Let us mention that the physical limit corresponding to a
vanishing elementary time step $\Delta \tau \to 0$ is trivially
taken in this algorithm. In this limit, the distribution of the escape
time $\tau$ from the configuration $\s$ is
\begin{equation}
P_\s(\tau)=\lambda\exp [-\lambda \tau],
\label{expocont}
\end{equation}
where $\lambda$
is related to the transition probabilities of (\ref{transition})
by $\lambda= \sum_m  p(\sigma \rightarrow \sigma^{[m]})/ \Delta\tau$.

In the above method, the sampling
of the dynamical evolution eq. (\ref{MCsequence}) is split into two
processes, the generation of the trajectories, and the sampling of
physical escape times with the probability distribution
$P_\s(\tau)$.
These two processes are {\it independent}
sources of statistical noise and the second one can be removed.
Consider a trajectory
$\sigma_1(\lambda_1) \rightarrow \sigma_2(\lambda_2)
\ldots \rightarrow \sigma_M(\lambda_M)$, where
$\lambda_i$ is the  {\it mean} escape time of configuration $\s_i$,
characterizing $P_{\sigma_i}(\tau)$ as in eq. (\ref{expocont}). For any
realization, the physical times have to be sampled from
eq. (\ref{expocont}).
The probability $P_M(\tau)$ to be at configuration $\s_M$ at a
given time $\tau$
is a convolution which
can be computed by the calculus of residues. In the special
case in which all the $\lambda_i$'s are different from each other, the
result takes the simple form:
\begin{equation}
P_M(\tau)=\sum_{j=1}^{M}\;\;\sum_{i=1 \neq j}^{M}\;\;\;\frac{\lambda_i}
{\lambda_i -\lambda_j}[1 - \exp(-\lambda_j \tau)]
\end{equation}
Unfortunately, we have not succeeded in  calculating
this quantity in a stable way for large $M$ (such as $M\sim 10^6
\sim 10^8$),
which is  the case of practical interest here.
A practical answer, and the one which we have
adopted, is simply to calculate, for a given trajectory eq.
 (\ref{MCsequence}),
a fair number of realizations
of the escape times (with a completely unsophisticated program,
we can calculate an average over $\sim 100$ realizations of the
escape times in just about twice the time it takes
to generate the trajectory).

We have used this algorithm to study   the LABS model.
We begin with  the results from simulated annealing. In fig.~1 we plot
 the internal energy per spin {\it vs} temperature at $N=100$.
The annealing
scheme starts from a temperature $T_1=.6$. After thermalizing at this
temperature,
the temperature is lowered down to $T_0=.04$ using the logarithmic scheme:
$T(t)=T_1- (T_1-T_0) Log(t)/Log(t_m)$, where the duration of the
run, $t_m$ (measured in Monte Carlo steps per Spin (MCS)),
takes the values $1.25^{20},1.5^{20},1.75^{20},2^{20},2.25^{20}, 2.5^{20}$.
So the logarithmic cooling rate $dT/d(ln(t))$ varies from
$.125$ to $ .0306$. The data
is averaged over $400$ to $50$ realizations of the random
 process of eq. (\ref{MCsequence}) with random initial conditions;
for each sequence of configurations we have studied a single realization
of the escape times. The system undergoes a freezing phenomenon at
a temperature which depends on the logarithmic cooling rate,
 reminiscent of the observations in glasses. An extrapolation of
the zero-temperature energy as a function of the
logarithmic cooling rate is compatible with a zero-temperature
energy per spin
of the order of $.07 \sim .08$. This result is about twice as
large as the
value $E_c=.0406$. $E_c$ is conjectured by the Bernasconi-Golay
approximation \cite{bernasconi,golay}, recently
derived using an approximate description of this system by one
with quenched disorder \cite{bm}, and is also compatible with
exact enumerations
of small size systems. We have done simulations at $N=200$, $N=400$
and $N=401$,
and $t_m= 1.25^{20},1.5^{20},1.75^{20},2^{20}$ which are nearly
indistinguishable from the data of fig. 1. The difference between
the result of simulated annealing and $E_c$ may well be due to
a dynamic freezing effect, which forbids to find the exact ground state
for large size systems.

We have performed a more detailed study of the dynamics
at temperatures $T=.075$, which is clearly in the low $T$ regime on all
the time scales we can reach, and at $T=.25$ for comparison.
We start by reviewing
the results at the lower temperature, $T=.075$.
In fig.~2 we plot the correlation function $C(t_w+\tau,t_w)= (1/N) \sum_i
<\s_i(t_w+\tau) \s_i(t_w)>$ {\it vs} the time $\tau$ for various
waiting times $t_w$, ranging up to $10^7$ (from now on all the times are
measured in
MCS). The aging effect is clear, on all the range of waiting times
we can reach.
The behavior is well approximated
 by a function of $\tau/t_w$, decreasing as a power law
$(\tau/t_w)^{-.026}$ at large arguments, as shown in fig.~3. This is
similar to what has been found in experiments \cite{ageexp},
and numerical simulations of spin glasses \cite{agesimu}
 or other disordered systems \cite{mapa}. Recent analytical studies of some
mean field spin glasses confirm the possibility of such a scaling
\cite{cuku} for a category of spin glasses characterized by a first order
(``one step'') replica symmetry breaking scheme like the Random
Energy Model \cite{derr,gm}, while the spin glasses
with full replica symmetry breaking seem to have
 a more complicated behavior
\cite{framez,cukusk}. We exhibit here a clear aging effect
 for the LABS model which is
a system without any quenched disorder. This data again suggests
that the LABS could be a good toy model for the glass transition.

On this aging data one also notices that the asymptotic dynamics
is essentially frozen: $lim_{\tau \to \infty} lim_{t_w \to \infty} C(t_w
+\tau,t_w) \simeq 1$.
It is interesting to look at
 a single run
 of a system with $N=400$ spins, at $T=0.075$. We show such a run,
typical of what we observe, in fig. 4.
 The simulation, which  has taken
a few hours on a HP 730 workstation, was stopped after a
time of $\tau=10^9$  MCS.
It is immediately seen on the plot of the energy {\it vs} time,
that the energy of the system is not equilibrating on this time scale.
 Clearly the system gets trapped into metastable states
with a long lifetime which are well identified by the plateaux in the
energy. The fluctuations along each plateau are quite small,
 which is in agreement
with the above observations: the metastable states at low temperature
 involve
a small number of
configurations. In contrast we show in figs 5 and 6 the analogous data
at a higher temperature $T=.25$. There is no trace of the plateaux in the
energy, and the system equilibrates after a finite
time $t \simeq 10^3$, which agrees with the simulated anealing data from
fig.~1. On the correlation function one observes an interrupted
aging effect.

The trapping behavior is reminiscent of the
simple generic model of aging proposed by Bouchaud
 \cite{bouchaud}. At a qualitative level
one does see on fig.~4 that the trapping times (length of the
plateaux) become longer when the waiting time increases.
A quantity which is easily accessible with our algorithm is the
distribution of escape time from single configurations
$\tilde \tau=1/\lambda$.
 (Notice that this is
not exactly the same as the trapping times because the plateaux involve
several configurations). In fig.~7 we plot
$\int_\tau^\infty P_k(\tau')
  \ d \tau'$ {\it vs} $\tau$,
where $P_k(\tau')$ is the distribution of escape time for the
$k$th visited
configuration in the sequence eq. (\ref{MCsequence}). We see a clear
increase of the
typical escape time with $k$ (and therefore with the age of the system).
For large waiting time (large $k$) the
 distribution $P_k(\tau')$ is broad, and compatible with the
idea of  \cite{bouchaud} that the average escape time is infinite.
The behavior of $P_k(\tau)$ at large times $\tau$ can be
approximated by a power law $P_k(\tau) \simeq \tau^{-1.6}$.
 The broadness of this distribution explains why the
algorithm we use is so efficient on this problem.
It  circumvents the problem of trapping by a single
configuration, but does not get rid of ``renormalized traps"
consisting of a group of configurations separated from the
rest by barriers. A cluster Monte Carlo algorithm \cite{kraplu}
has been developed to handle this problem.

To conclude, we have found a model without quenched disorder which
is another representative of a well known class of glassy systems,
those which
undergo a "one step replica symmetry breaking" transition in the
static approach. Other members of this class
include the Random Energy Model \cite{derr,gm},
the binary perceptron \cite{kramez}, the p-spin spherical spin glass
\cite{kirthi}, random heteropolymers
\cite{garorl,shagut,iormarpar},
the manifolds in random media with short range disorder
\cite{mezpar}, {\sl etc}. We believe that the algorithm we have described
will be efficient for any of these problems. Our code is available
by e-mail.

\acknowledgments
We acknowledge helpful discussions with J.P. Bouchaud, E. Marinari
and G. Parisi.
\\

\section*{Figure Captions}
\begin{description}

\item{\bf Fig. 1:}
\noindent
Simulated annealing results of the energy per spin {\it vs}
temperature in a $N=100$ spin system,
corresponding to  logarithmic cooling rates (from top to bottom)
$.125, .07, .05, .04, .034, .03$.

\item{\bf Fig. 2:}
\noindent
The two times correlation of a $N=400$
spin system at temperature $T=.075$,
 $C(t_w+\tau,t_w)$ {\it vs} $\tau$, plotted for
various values of the waiting time $t_w=5^n$ with
$n=0,1,...,10$.
The data is averaged over 300 initial conditions (and for
each of them 50 realizations of the sampling of escape times).

\item{\bf Fig. 3:}
\noindent
The same data as fig.2 plotted {\it vs} $\tau/t_w$, for
$t_w=5^{2n}$ with
$n=0,1,...,5$.

\item{\bf Fig. 4:}
\noindent
Energy per spin {\it vs} $\log_{10}$(time)
 in a single run for a $N=400$ spin system at $T=.075$.

\item{\bf Fig. 5:}
\noindent
The two times correlation of a $N=400$
spin system at temperature $T=.25$,
 $C(t_w+\tau,t_w)$ {\it vs} $\tau$, plotted for
various values of the waiting time $t_w=2.5^n$ with
$n=0,1,...,7$.
The data is averaged over 400 initial conditions (and for
each of them 50 realizations of the sampling of escape times).

\item{\bf Fig. 6:}
\noindent
Energy per spin {\it vs} $\log_{10}$(time)
 in a single run for a $N=400$ spin system at $T=.25$.

\item{\bf Fig. 7:}
\noindent
The integrated distribution of escape times, $\int_\tau^\infty P_k(\tau')
  \ d \tau'$, for the $k$th  visited configuration, is
plotted
{\it vs} $\tau$ for $k=4^n$ with $n=2,...,7$, in a simulation of
a $N=400$ spin system  at $T=.075$.

\end{description}

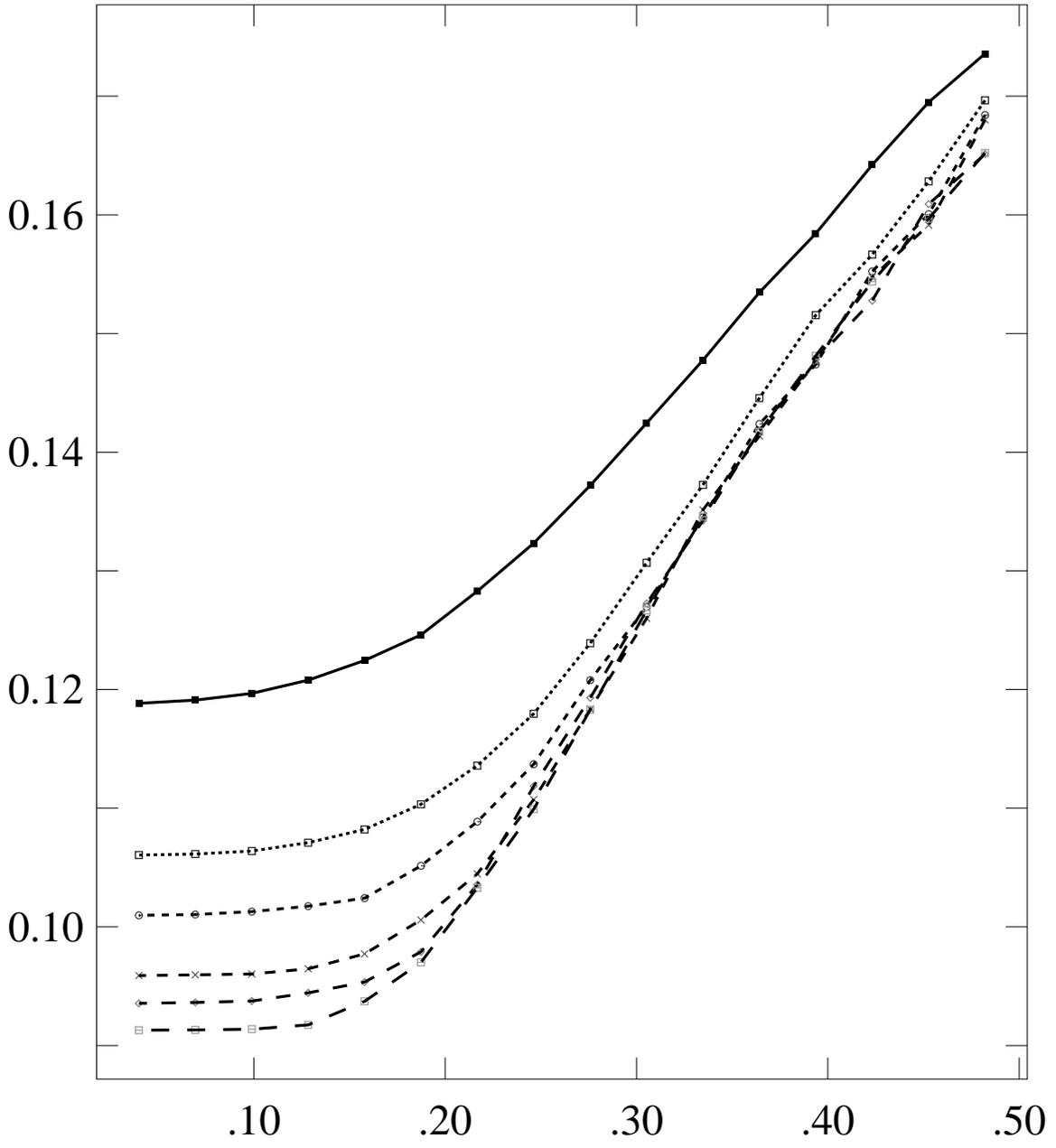
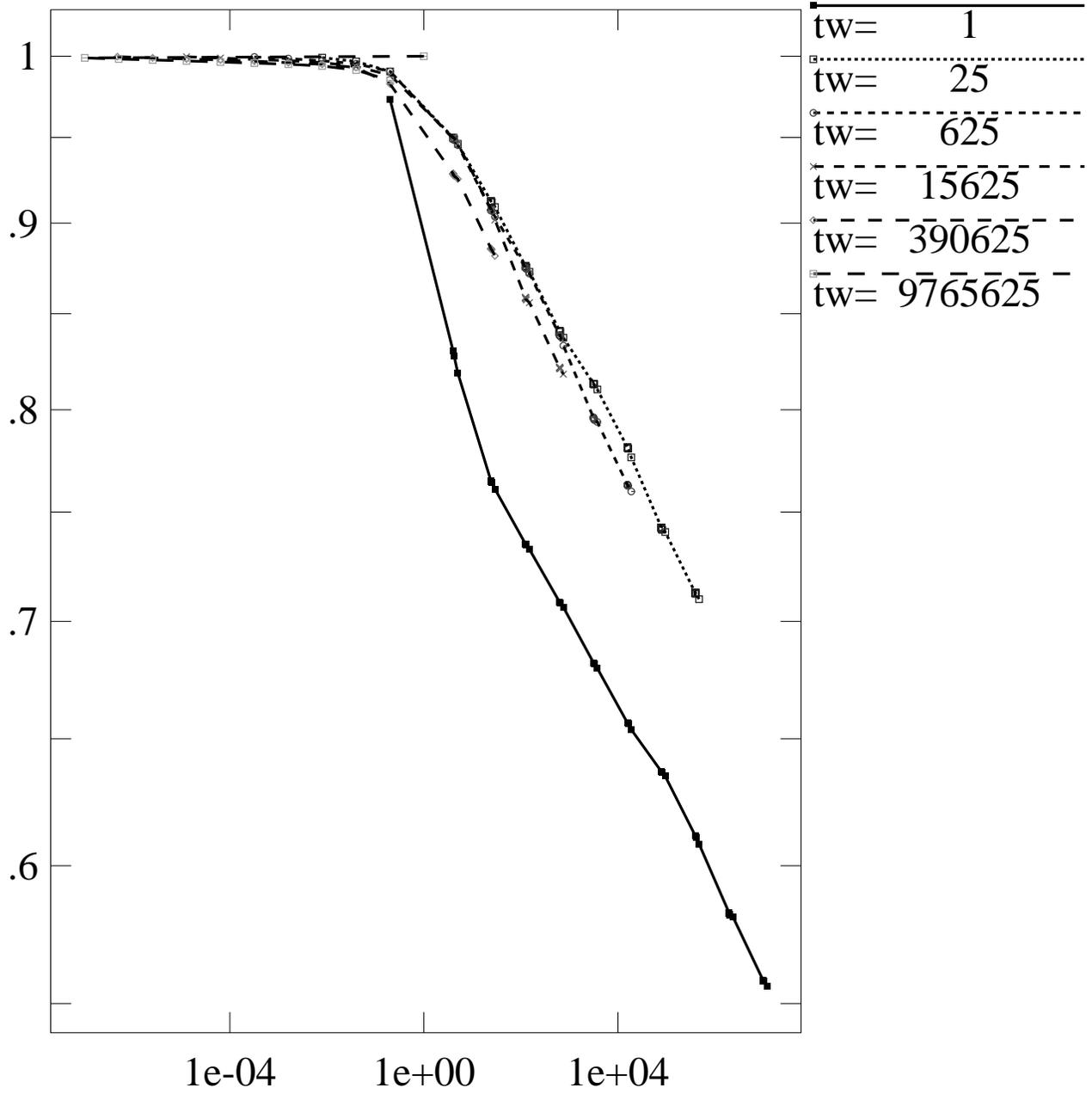
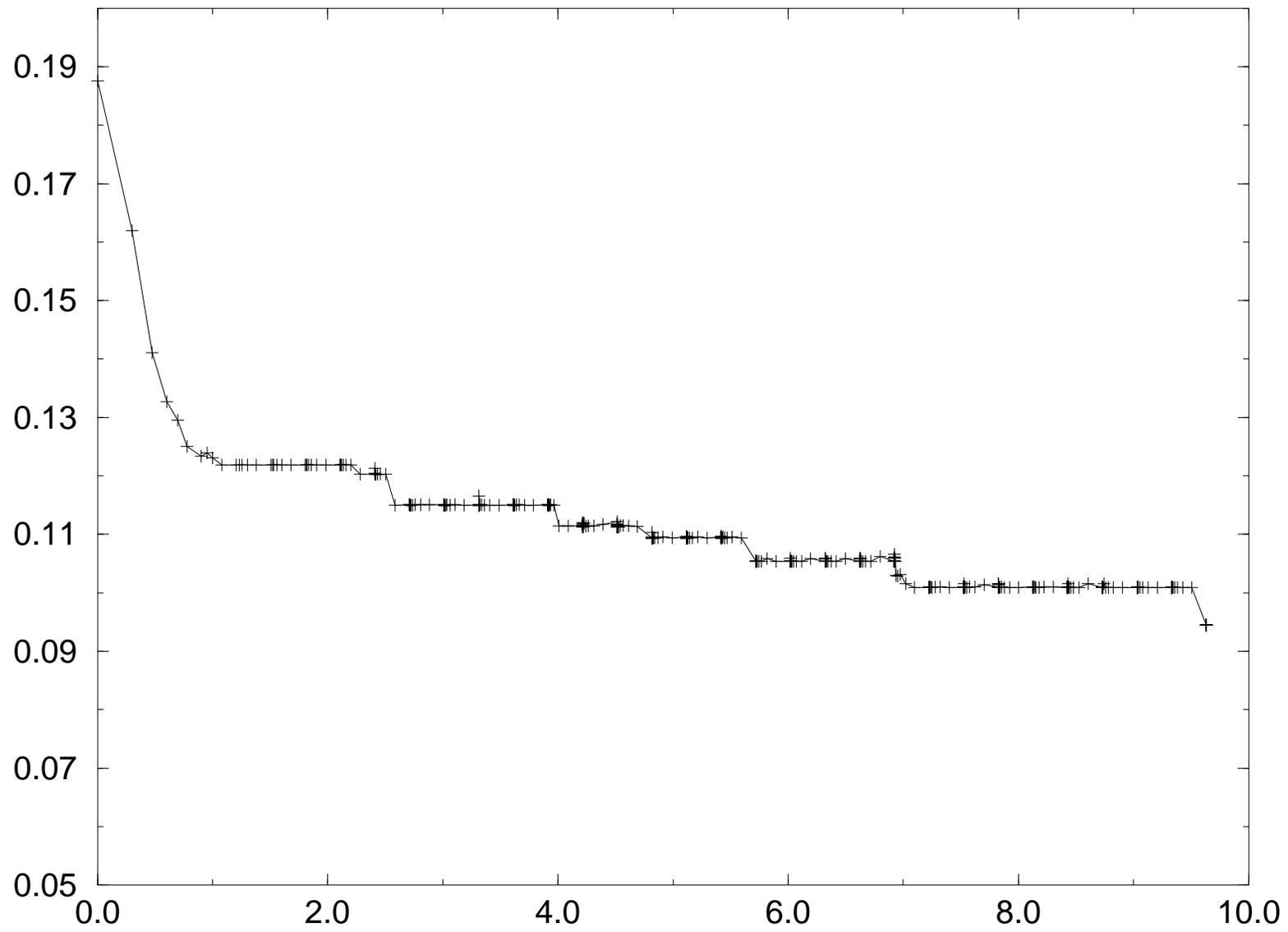
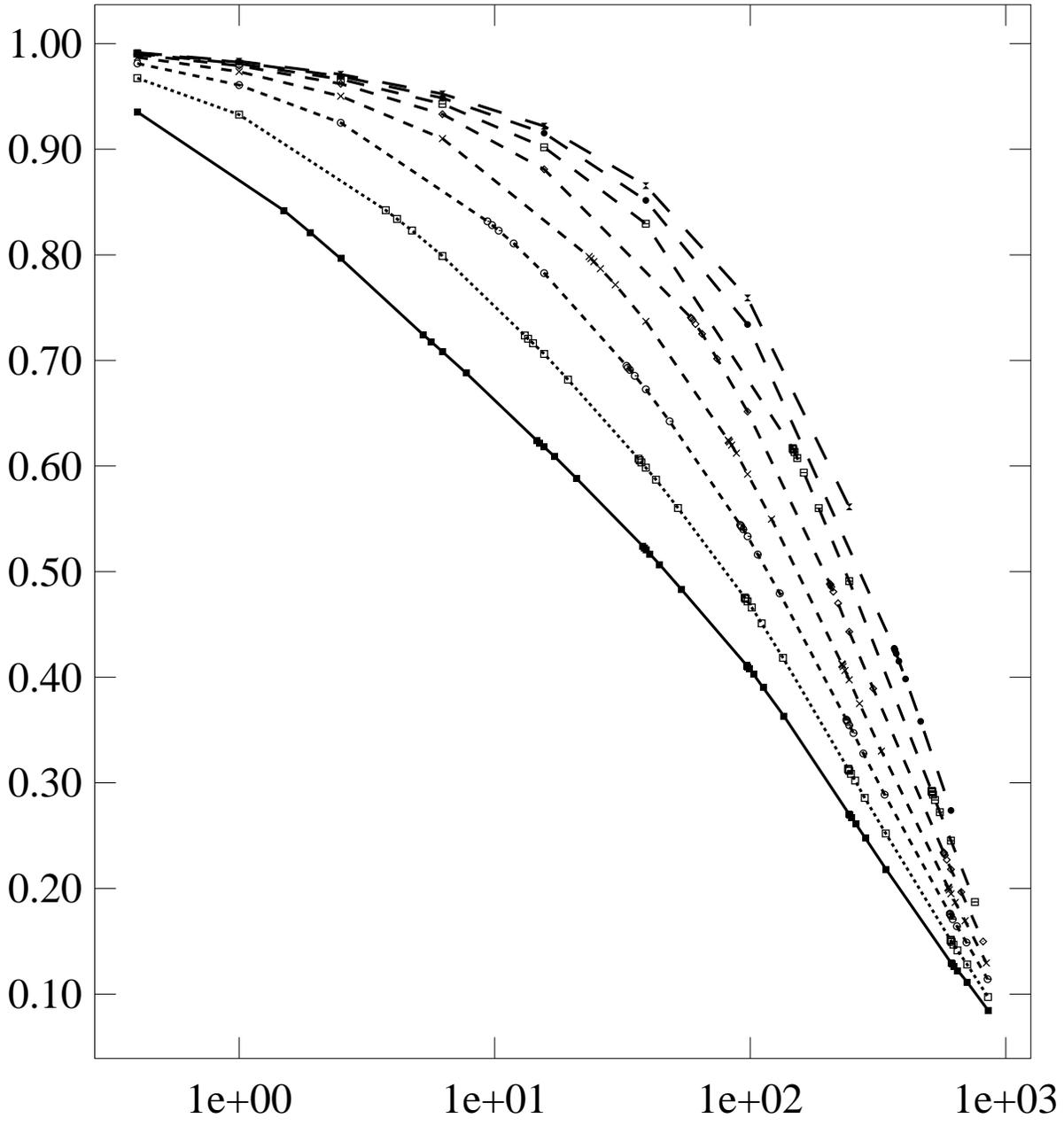
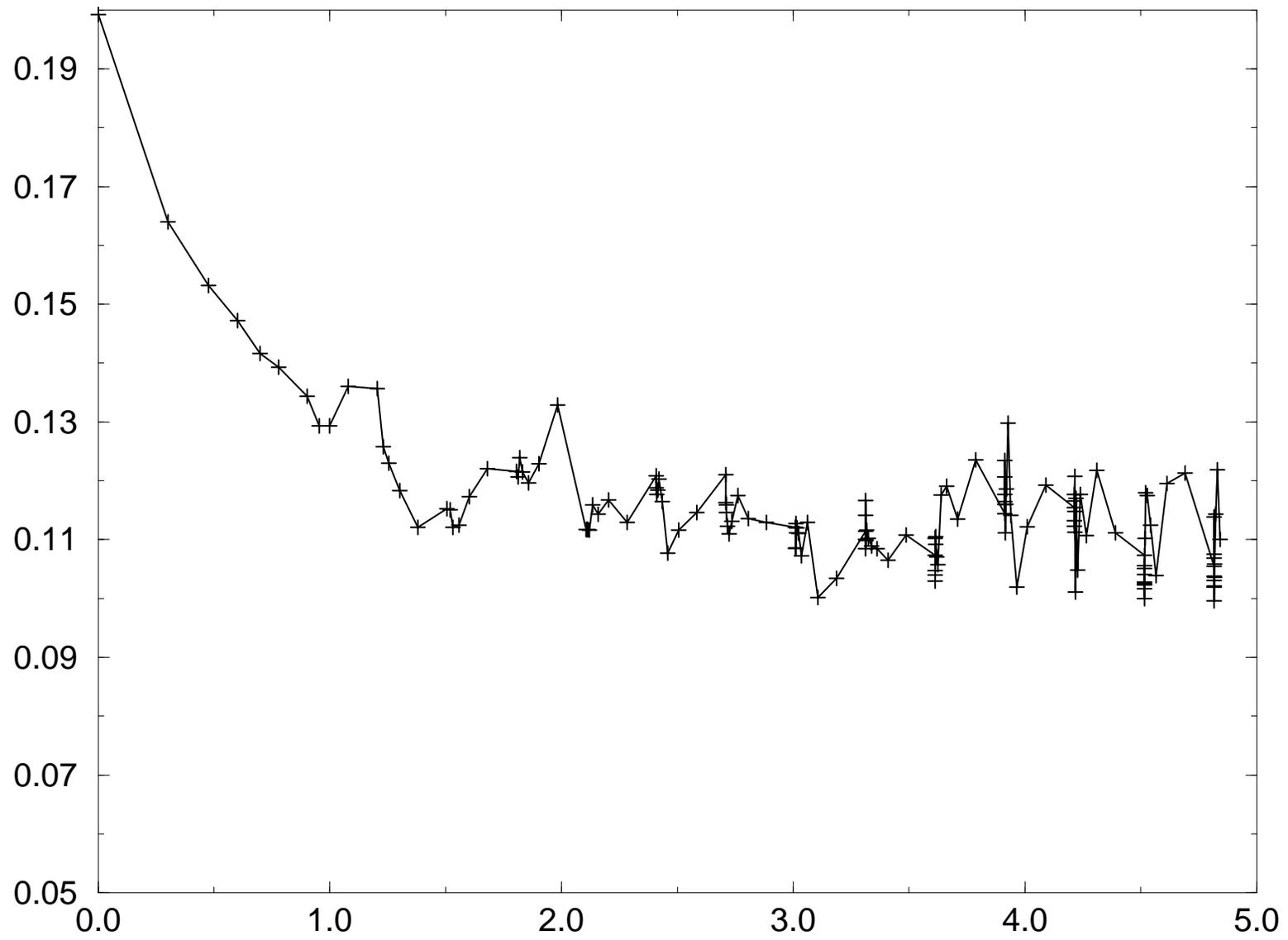
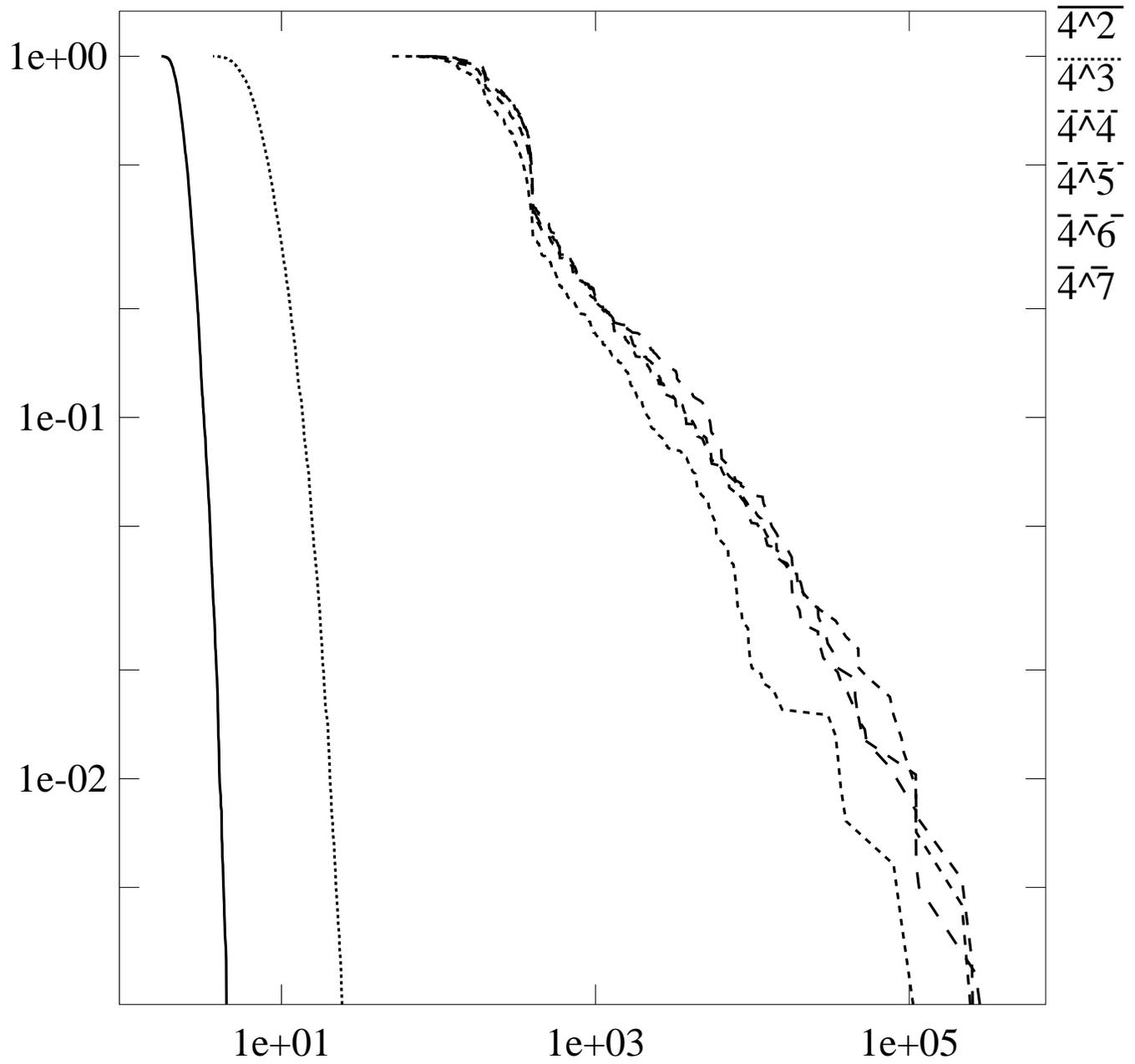
\end{document}